\documentclass[aps,prb,twocolumn,groupedaddress,showpacs,fleqn,floatfix]{revtex4}

\usepackage{amsmath}
\usepackage{epsfig}
\usepackage{color}
\usepackage{graphics}

\newcommand{\figa} {
\begin{figure}
\centering
\includegraphics[width=4cm,clip]{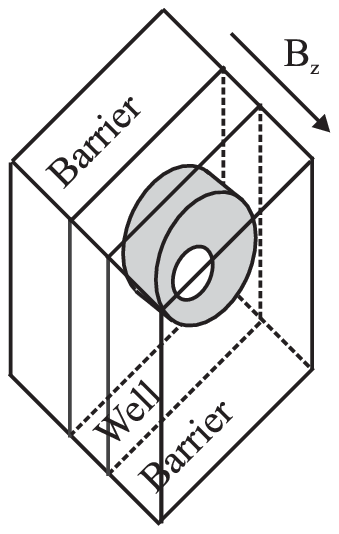}
\caption{Schematic drawing of the investigated non-circular (eccentric)
nanoring (shaded) embedded in a quantum well} \label{figa}
\end{figure}
}

\newcommand{\figb} {
\begin{figure}
\centering
\includegraphics[width=8cm,clip]{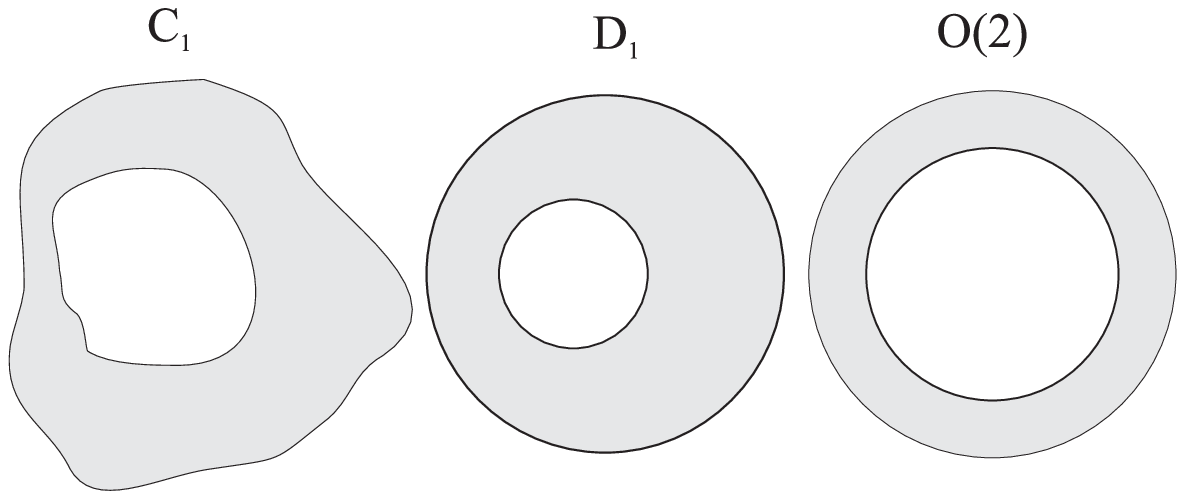}
\caption{Schematic picture of nanorings with different
symmetries.} \label{figb}
\end{figure}
}

\newcommand{\figc} {
\begin{figure}
\centering
\includegraphics[width=5cm,clip]{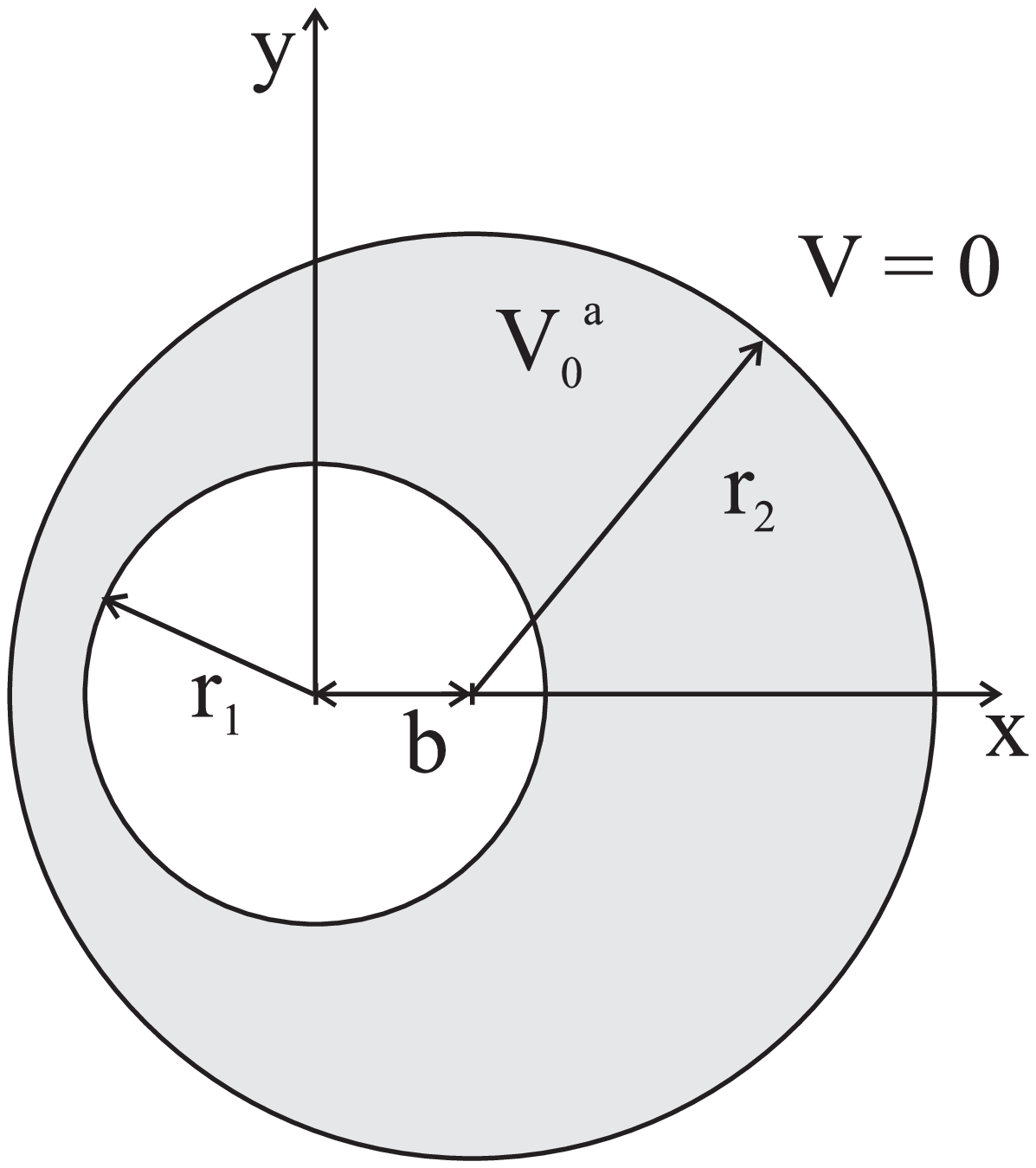}
\caption{Schematic picture of a nanoring with $D_1$ symmetry.
Within the gray region, the potential is nonzero. Inner and outer
ring boundaries are circles with radii $r_1$ and $r_2$, the
centers of which are displaced by $b$.} \label{figc}
\end{figure}
}

\newcommand{\figca} {
\begin{figure}
\includegraphics[width=8cm,clip]{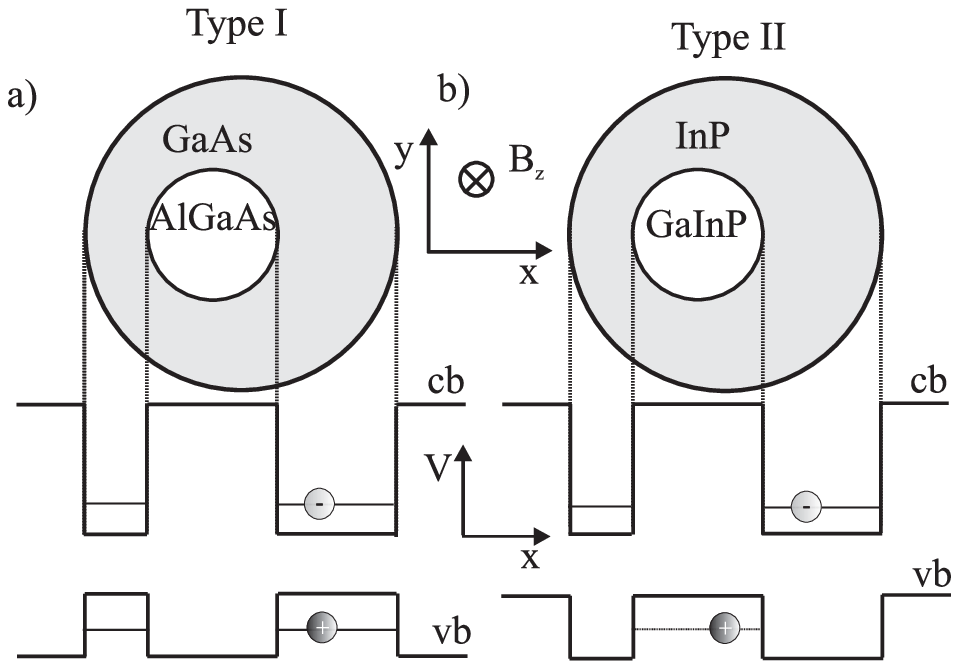}
\caption{Schematic view of in-plane geometry (top) and energy
profiles for conduction and valence band (bottom) for the
investigated nanorings of type I a) and type II b). Specific
electron and hole positions are visualized.} \label{figca}
\end{figure}
}

\newcommand{\figd}  {
\begin{figure*}
\centering
\includegraphics[width=11cm,clip]{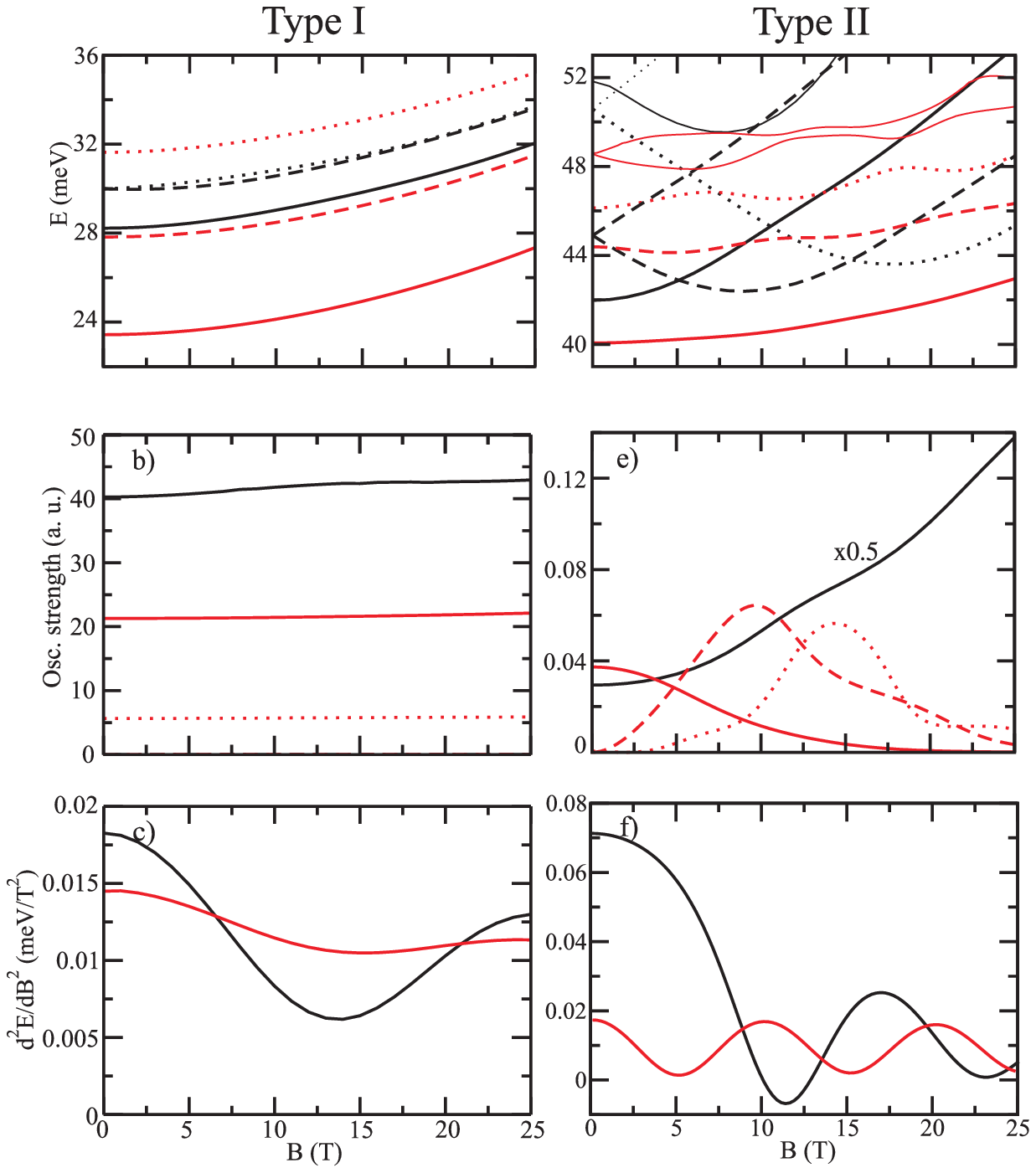}
\caption{(Color online) Calculated exciton properties in
nanorings in dependence on magnetic field. The left column refers
to a type I system (GaAs/AlGaAs) with ring radii $r_1= 4$~nm,
$r_2 = 12$~nm, while for the right column a type II system
(InP/GaInP) with $r_1= 8\,$nm and $r_2 = 16\,$nm has been
considered. Results for a circular ring ($b = 0$) are shown in
black, while red (grey) curves are obtained for a non-circular ring (type I: $b
= 1$~nm, type II: $b = 0.5$~nm). The lowest exciton levels are shown in a) and d) (for details see text). The spin-dependent Zeeman energy is not included. In b) and c) the corresponding oscillator
strengths $f_\alpha^2$ are plotted. The lowest panel c) and d) show the second
derivative of the ground state energy which quantifies the X-ABE
oscillations.} \label{figd}
\end{figure*}
}

\newcommand{\figf}
{
\begin{figure}
\centering
\includegraphics[width=8cm,clip]{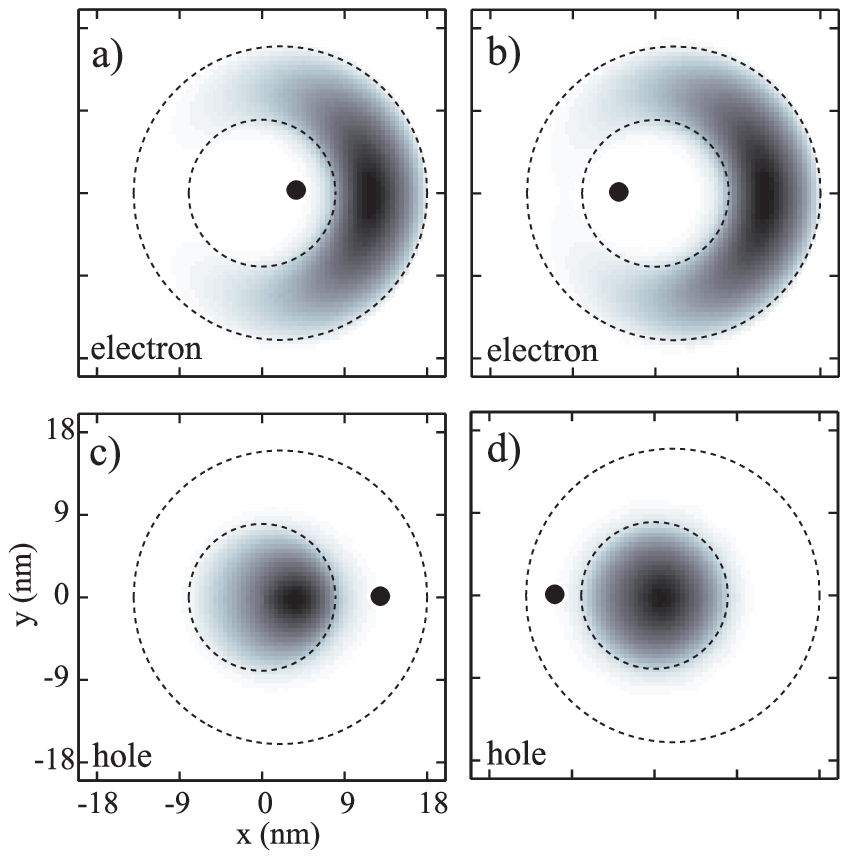}
\caption{Correlated electron and hole densities of the exciton
ground state at zero magnetic field for a non-circular type II
nanoring (InP/GaInP, $r_1 = 8\,$nm, $r_2 = 16\,$nm, $b=2\,$nm).
The black dot indicates the chosen coordinates of the other particle.}
\label{figf}
\end{figure}
}

\newcommand{\figi}  {
\begin{figure*}
\centering
\includegraphics[width=7cm,clip,angle=-0]{Fig07a.eps}
\includegraphics[width=7cm,clip,angle=-0]{Fig07b.eps}
\caption{(Color online) Calculated exciton (solid) and free electron-hole pair (dashed) properties in non-circular nanorings of type II (InP/GaInP) with $r_1= 8\,$nm, $r_2 = 16\,$nm, and $b= 2$~nm as a function of magnetic field. The ground state in black, the first and the second excited state in red (dark grey) and green (light grey) are shown in a) (the spin-dependent Zeeman energy is not included). In b) the second derivative of the ground state energy is depicted.} \label{figi}
\end{figure*}
}

\newcommand{\figga} {
\begin{figure}
\centering
\includegraphics[width=8cm,clip]{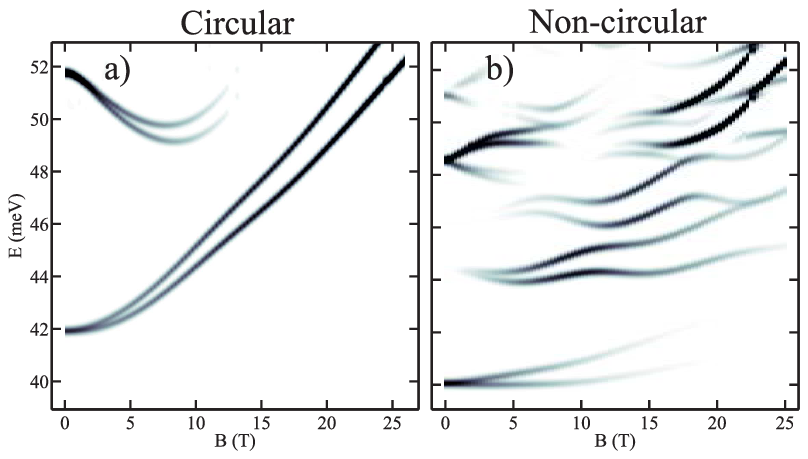}
\caption{Calculated absorption spectra of a circular ($b=0\,$nm) and non-circular ($b=0.5\,$nm) type II nanoring (InP/GaInP, $r_1 = 8\,$nm, $r_2 = 16\,$nm) including the spin contribution Eq.~(\ref{spins}). All spectra are Gauss broadened with variance $\sigma=0.1\,$meV and displayed using a linear gray scale. } 
\label{figga}
\end{figure}
}

\newcommand{\figgb} {
\begin{figure}
\centering
\includegraphics[width=8cm,clip]{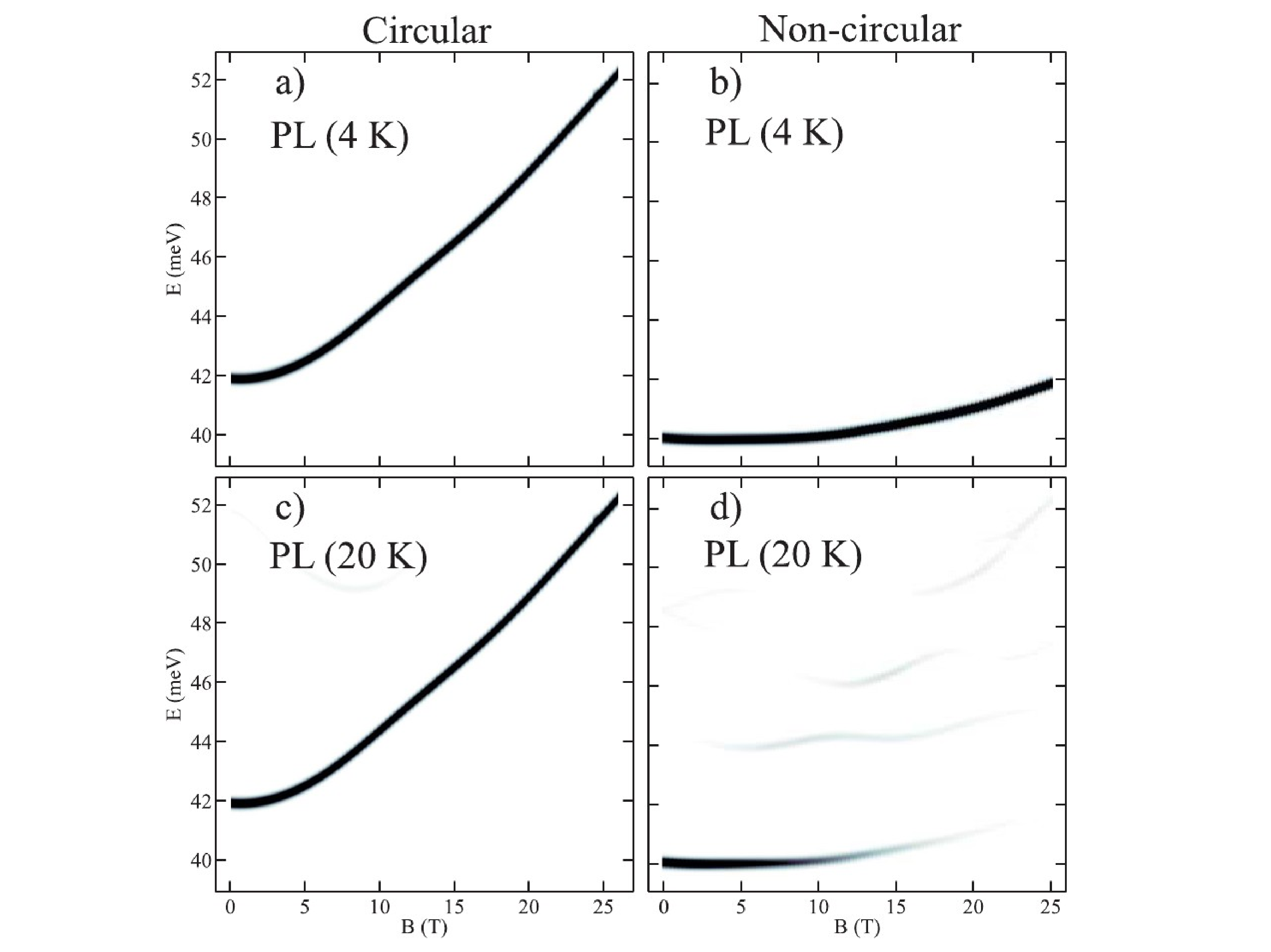}
\caption{Calculated photoluminescence spectra of a circular (a and c) and a non-circular (b and d) type II nanoring as in Fig.~\ref{figga}. Exciton occupations are calculated from the kinetic equation Eq.~(\ref{kinet}), fixing the phonon temperatures as indicated. Without non-radiative decay, $d_\alpha=0$.} 
\label{figgb}
\end{figure}
}

\newcommand{\figgc} {
\begin{figure}
\centering
\includegraphics[width=8cm,clip]{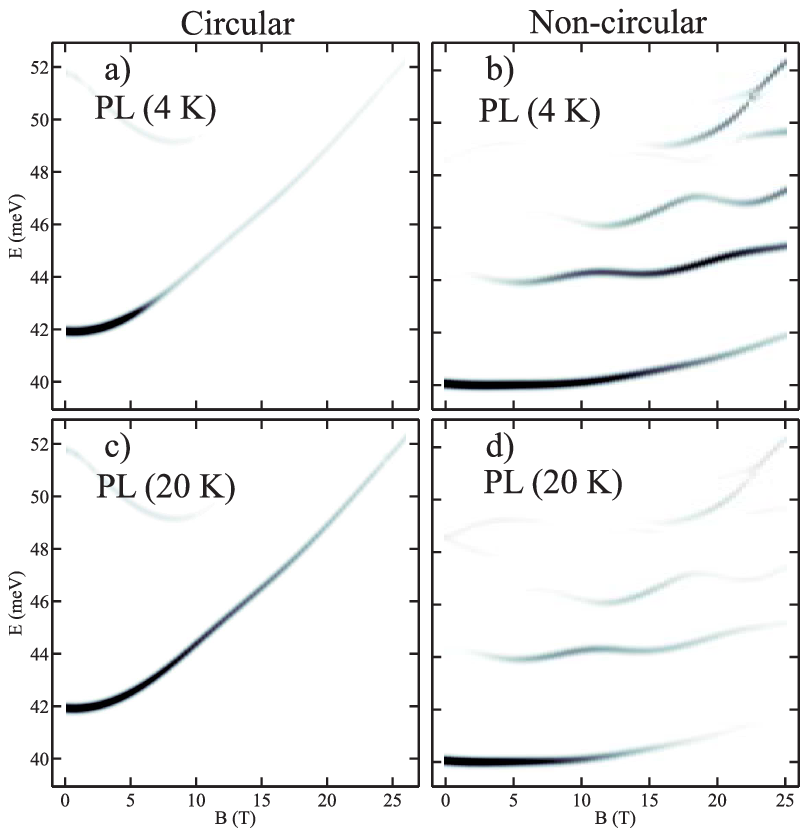}
\caption{Calculated photoluminescence spectra as in Fig.~\ref{figgb}, but here with a state independent non-radiative rate of $d_\alpha = 10\,$ns$^{-1}$.} 
\label{figgc}
\end{figure}
}

\newcommand{\figh} {
\begin{figure*}
\centering
\includegraphics[width=11cm,clip]{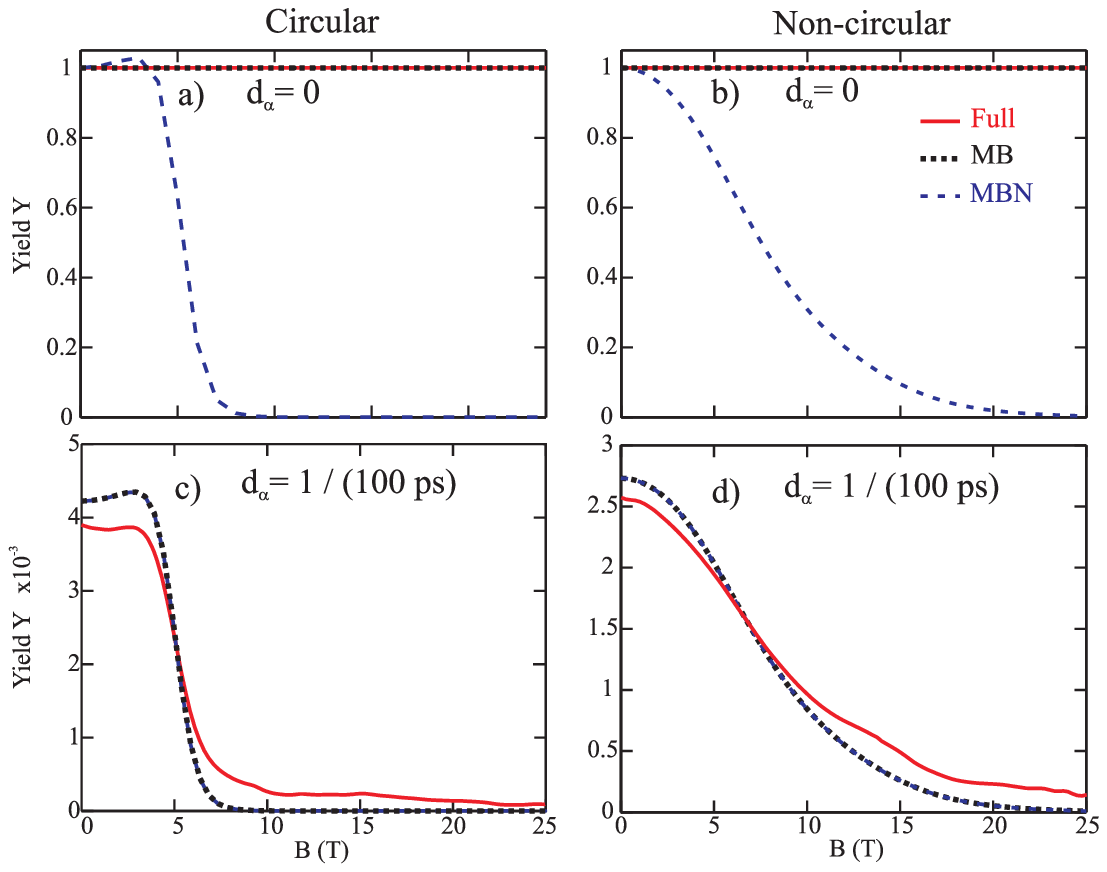}
\caption{(Color online) The photoluminescence yield $Y$ for the circular (a and c) and non-circular (b and d) nanorings of
Fig.~\ref{figga} at $T = 4\,$K and two values of the non-radiative
decay rate $d_\alpha = 0$ a) and $d_\alpha = 10\,$ns$^{-1}$ b).
The full solution Eq.~(\ref{full}) (Full, solid) is compared to
the assumption of a Maxwell-Boltzmann distribution with constant
pump rate Eq.~(\ref{MB}) (MB, dotted)  and with constant exciton
density Eq.~(\ref{govorov}) (MBN, dashed) is plotted.}
\label{figh}
\end{figure*}
}

\newcommand{\tabc}
{
\begin{table}
\centering \caption{The symmetry of the ring geometry leads to
restrictions for the matrix elements of the in-plane potential
Eq.~(\ref{matel}).} \label{tabc}
\begin{ruledtabular}
\begin{tabular}{cl}
 Point group of the ring & Non-zero elements of $V^a_{k}$ at\\
 \hline
$C_n$ & $k = j n$\, ($j$ integer)\\
$D_n$ & $k = j n$\, ($j$ integer) and $V^a_k = V^a_{-k}$\\
$O(2)$ & $k = 0$\\
\end{tabular}
\end{ruledtabular}
\end{table}
}

\newcommand{\taba}
{
\begin{table}
\caption{Confinement potentials $V^a_0$, static dielectric
constant $\epsilon_S$, in-plane effective masses $m_a$, $g$
factors $g^*_a$, mass density $\rho_M$, sound velocity $s$, and
deformation potentials for conduction ($D_c$) and valence band
($D_v$).} \label{taba}
\begin{ruledtabular}
\begin{tabular}{ccc}
 & GaAs/Al$_{0.23}$Ga$_{0.77}$As & InP/Ga$_{0.51}$In$_{0.49}$P \\
\hline
$V^e_0$ (meV) & -257 \footnotemark[3]& -600 \footnotemark[2] \\
$V^h_0$ (meV) & -110 \footnotemark[3]& 50 \footnotemark[2]\\
$\epsilon_S$ & 12.5 \footnotemark[1]& 12.6 \footnotemark[2]\\
$m_e/m_0$ & 0.067 \footnotemark[1]& 0.077 \footnotemark[2]\\
$m_h/m_0$ & 0.36 \footnotemark[1]& 0.6 \footnotemark[4] \\
$g^*_e$ & 0.1 \footnotemark[7] & 1.6 \footnotemark[8] \\
$g^*_h$ & -1.2 \footnotemark[7] & -3.0 \footnotemark[8] \\
$\rho_M$ (kg m$^{-3}$) & 5370 \footnotemark[6] & 4810 \footnotemark[5] \\
$s$ (m/s) & 5330 \footnotemark[6]& 5230 \footnotemark[5] \\
$D_c$ (eV) & 7.0 \footnotemark[6]& 6.0 \footnotemark[3]\\
$D_v$ (eV )& -3.5 \footnotemark[6]& -0.6 \footnotemark[3]\\
\end{tabular}
\end{ruledtabular}
\footnotetext[1]{Ref.~\onlinecite{SAK2001}.}~\footnotetext[2]{Ref.~\onlinecite{TPJ+2002}.}
\footnotetext[3]{Ref.~\onlinecite{VMR2001}.}
\footnotetext[4]{Ref.~\onlinecite{JPP2001}.}
\footnotetext[5]{Ref.~\onlinecite{Adachi92}.}
\footnotetext[6]{Ref.~\onlinecite{SAK2001}.}
\footnotetext[7]{Ref.~\onlinecite{SBM+1992}.}
\footnotetext[8]{Ref.~\onlinecite{SRK+1996}.}
\end{table}
}

\begin{document}

\title{Non-circular semiconductor nanorings of type I and II: \\
Emission kinetics in the exciton Aharonov-Bohm effect}

\author{Michal Grochol and Roland Zimmermann}

\affiliation{Institut f\"{u}r Physik der Humboldt-Universit\"{a}t
zu Berlin, Newtonstr. 15, 12489 Berlin, Germany}

\date{\today}

\begin{abstract}
Transition energies and oscillator strengths of excitons in dependence on magnetic field are investigated in type I and II semiconductor nanorings. A slight deviation from circular (concentric) shape of the type II nanoring gives a better observability of the Aharonov-Bohm oscillations since the ground state is always optically active. Kinetic equations for the exciton occupation are solved with acoustic phonon scattering as the major relaxation process, and absorption and luminescence spectra are calculated showing deviations from equilibrium. The presence of a non-radiative exciton decay leads to a quenching of the
integrated photoluminescence with magnetic field.
\end{abstract}


\maketitle

\section{Introduction}

The Aharonov-Bohm effect (ABE) rests upon the action of the
vector potential on quantum mechanical particles. The idea behind
is quite simple: A charge particle orbiting around a region with
non-zero magnetic flux $\Phi_B = \int {\bf B} \cdot d{\bf S}$
aquires an energy which is a periodic function of the magnetic
flux $\Phi_B$, the period being given by the magnetic flux
quantum $h / e$. Shortly after its theoretical
prediction,\cite{AB59} the ABE has been observed experimentally.\cite{Ch1960, MB1962} Doped semiconductor nanorings
are among actual realizations\cite{LLG+2000} where the ABE could
be detected with high precision.

The exciton Aharonov-Bohm effect is a more recent invention: The
neutral exciton consisting of electron and hole is predicted to
show a similar oscillating behavior. Originally, a related
many-body system has been studied by Wendler and
coworkers\cite{WF1994, WFC+1996} who have considered the Coulomb
effect when placing two electrons into a quantum ring -- instead
of electron and hole as in the present exciton case. They form a
kind of Wigner molecule with interesting properties like
crystallization. The approximations used originally allowed for
an analytic solution of the ground state and the related
persistent current. Later on, a full calculation was carried out,
but always using a rigid lateral confinement in the
ring.\cite{PSN2004}

For the exciton Aharonov-Bohm effect (X-ABE), the limiting case
of a nanoring with zero width has been studied first as
well.\cite{Ch1995, RR2000} The simplicity of the model allowed to
investigate many aspects of X-ABE like the Berry phase and
persistent current,\cite{MC2004} absorption,\cite{CM2005} or
deviation from circular symmetry due to the presence of
impurities.\cite{SUG2004, SUS2005} One basic result was that the
X-ABE can only be observed as long as the ring diameter is
comparable or below the exciton Bohr radius $a_B$.~\cite{RR2000}
Since the ring size cannot be made arbitrarily small, a better
idea is to increase $a_B$, i.e. to weaken the exciton. This can
be achieved by separating electron and hole using a static
electric field,\cite{MC2003, ZZC2005} or going to type II
material combinations where electron and hole are confined in
different regions.\cite{GUKW2002, SUG2004, SUS2005, BPS+2006}

In the original ABE idea, the particle path was assumed to lie in
a region with zero magnetic field, which means to concentrate the
magnetic flux into the middle of the ring. For a semiconductor
nanoring of nanometer size, this is technically very demanding --
if not impossible using state-of-the-art techniques. Therefore,
for X-ABE a situation with the magnetic field penetrating
homogeneously the nanoring is more realistic. Then, the energy in
dependence on magnetic field acquires a non-oscillating part
roughly proportional to $B^2$ as well which can complicate the
extraction of X-ABE oscillations enormously.

First calculations for excitons in a {\it circular nanoring} with
{\it finite width} and homogeneous penetrating magnetic
field\cite{HZLX2001,SU2001, GBW2002} could not find the X-ABE for
the ground state. Theoretical progress was made by calculations
on two-dimensional annular lattices.\cite{PDER2005} X-ABE
oscillations were found but the $B^2$ energy shift was not
included. Our recent calculations for type I and II nanorings
with finite width in Ref.~\onlinecite{GGZ2006b} have clearly
shown that the ground state energy has an oscillatory component,
which is related to the exciton persistent current. For
extracting its amplitude, we have proposed to calculate the
second derivative of the energy with respect to $B$. On the
search for optimal nanoring parameters, we have discussed in
detail different material combinations for type II nanorings.

Compared to Ref.~\onlinecite{GGZ2006b}, the present work deals with
{\it non-circular geometry} and gives a detailed
study of {\it exciton kinetics} in order to calculate exciton occupation which is not necessarily in
equilibrium with the lattice temperature. We restrict ourselves
to incoherent and steady-state excitation following
Ref.~\onlinecite{ZR1997}. A first attempt to calculate the
exciton photoluminescence (PL) in zero width nanorings with a single 
impurity (and also quantum dots with respect to X-ABE) has been
presented recently in Ref.~\onlinecite{SUG2004}. As a
simplification, however, an equilibrium Maxwell-Boltzmann
distribution for the excitons has been assumed, and only the
integrated PL intensity was shown. We have started with a more
realistic exciton kinetics, still in the zero width model for
circular nanorings in Ref.~\onlinecite{GGZ2006c}. Non-radiative
decay channels have been identified to be responsible for
the decrease of the integrated PL signal with increasing magnetic
field, which is named PL quenching.

The experimental situation for the X-ABE is far from clear: An
ensemble of InP/GaAs type II quantum {\it dots} has been studied
in Ref.~\onlinecite{RGC2004} and a theoretical explanation based
on~Ref.~\onlinecite{KKG1998} indicated some X-ABE oscillations in
a single dot. Later on, the non-circular shape of the quantum dot
($D_1$ and $D_2$ symmetry [see Section II]) has been taken into
account.~\cite{SUG2004} Furthermore, in a recent single dot
experiment~\cite{GGN+2006} on InP/GaAs quantum dots (grown under
different conditions) no oscillations have been observed. This is
consistent with our recent calculation for embedded InP/GaAs
quantum dots in Ref.~\onlinecite{GGZ2006a}. In a nanoring, only
the ABE for negatively charged excitons -- trions -- has been
observed.~\cite{BKH+2003} The X-ABE in nanorings still
waits for its unambigous experimental verification.
\figa

The paper is organized as follows. The exciton Hamiltonian is
given in Section II in radial and angular variables, and its
matrix elements are analyzed according to the ring symmetry
using arguments of group theory. Examples for energies,
oscillator strengths, and oscillation amplitudes of the X-ABE are
shown in Section III, for both type I and type II structures.
Kinetic equations for the exciton occupation including acoustic
phonon scattering and several decay channels are discussed in
Section IV. Its numerical solution allows to plot absorption and
photoluminescence spectra for various parameters (Section V).
The paper is concluded in Section VI, while more technical
details are deferred to the Appendix.

\section{Exciton Hamiltonian}

For the heavy-hole exciton in nanostructures, effective mass
theory with appropriate in-plane ($m_a$) and growth direction
masses ($m_{a,z}$) is used often. The subscript denotes electron
($a = e$) and hole quantities ($a = h$). Within this
approximation, we will investigate a structure which consists of
a narrow quantum well with confining potentials $U_a(z_a)$ into
which a nanoring of general symmetry is embedded. The
corresponding lateral confinement is given by $V_a({\bf r}_a)$.
The three-dimensional vector ${\mathsf r}$ is decomposed into its
two-dimensional in-plane part and the $z$-component in growth
direction, ${\mathsf r}_a = ({\bf r}_a, z_a)$. The ring structure
is schematically plotted in Fig.~\ref{figa}. Including a constant
$B$-field in $z$-direction (perpendicular to the quantum well),
we have the single exciton Hamiltonian
\begin{widetext}
\begin{eqnarray}
\label{H01} \hat{H} = \sum_{a=e,h} \biggl ( \frac{1}{2m_a}
\left (\hat{{\bf p}}_{{\bf r}_a} \mp e{\bf A}({\mathsf r}_a) \right)^2 +
\frac{1}{2m_{a, z}} \hat{p}^2_{z_a} + U_a(z_a) + V_a({\bf r}_a)
\pm g^*_a \mu_{B}B \sigma^z_a \biggr ) -\frac{e^2}{4\pi
\epsilon_0 \epsilon_S |{\mathsf r}_e - {\mathsf r}_h |} \, ,
\end{eqnarray}
\end{widetext}
where upper (lower) sign refer to electron (hole). $g^*_a$ are
effective $g$ factors, $\mu_{B} = e \hbar / 2 m_0$ is the Bohr
magneton, and $\sigma_a^z$ the Pauli spin matrix (along $z$). The
vector potential is used in symmetric gauge,  ${\bf A} ({\mathsf r})
= \frac{1}{2} {\bf B} \times {\mathsf r}$. The Coulomb potential
between electron and hole (last term) is screened by the static
dielectric constant $\epsilon_S$ of the semiconductor material.

Angular momentum and spin parts of the wave function can be
separated for the heavy-hole exciton, forming a quadruplet with the $z$-projection of the
total angular momentum $M=\pm 1$ (optically active or bright
states, with circular polarization $\sigma^\pm$) and $M = \pm 2$
(dark states). Combining electron and hole $g$ factors into
effective exciton $g$ factors\cite{SBM+1992, BSH+1994} as
$g^*_{X,\pm 1} = \pm(g^*_h+g^*_e)$ and $g^*_{X,\pm 2} =
\pm(g^*_h-g^*_e)$ allows to write the spin-dependent contribution
to the exciton energy as
\begin{eqnarray}
\label{spins} 
E^{spin}_M = \frac{1}{2}g^*_{X,M} \,\mu_{B}\,B \, ,
\end{eqnarray}
which gives rise to the Zeeman splitting linear in $B$.
In the following this spin contribution is not written explicitly
since its addition to the exciton energies is straightforward.
The exciton exchange interaction gives rise to an additional fine
structure splitting which is neglected here in view of the
dominant $B$-field effects.

Due to the confinement strength in $z$-direction, we factorize from the total wave function the product of sublevel wave functions \cite{Bastard} of the lowest electron and hole states in the quantum well. The remaining in-plane part of the wave function has to be calculated from the $z$-averaged Hamiltonian.\cite{GGZ2006b} Introducing polar coordinates and difference and sum angle [$ \phi = \phi_e - \phi_h$, $\Phi = (\phi_e + \phi_h)/2 $], we arrive at 
\begin{eqnarray}
\label{Ham03}
\nonumber \hat{H} &=& \sum_{a = e,h} \biggl[ -\frac{\hbar^2}{2m_a}
\frac{1}{r_a} \frac{\partial}{\partial r_a} \left (r_a \frac{\partial}{\partial r_a} \right) \\
&+& \frac{1}{2 m_a r_a^2}
\left ( i \hbar \left [ \frac{\partial}{\partial \phi} \pm \frac{1}{2} \frac{\partial}{\partial \Phi} \right ]
- \frac{eB}{2} r_a^2 \right)^2  \nonumber \\
&+& V_a(r_a, \Phi \pm \frac{1}{2} \phi) \biggr] - V_C(r_e, r_h,
\phi),
\end{eqnarray}
where $+$ ($-$) sign refers refers to electron (hole). The confinement averaged Coulomb potential $V_C(r_e, r_h, \phi)$ has been defined in Ref. \onlinecite{GGZ2006b}.

The exciton eigenfunction of the state $\alpha$ as solution of
\begin{eqnarray}
\hat{H} \Psi_\alpha = E_\alpha \Psi_\alpha
\end{eqnarray}
is expanded into the basis of angular momentum eigenfunctions,
\begin{eqnarray}
\label{expanG2} \Psi_\alpha(r_e, r_h, \phi, \Phi) = \frac{1}{2 \pi} \sum_{l, L}
u_{l, L, \alpha}(r_e, r_h) e^{il\phi} e^{i L \Phi}
\, .
\end{eqnarray}
The azimuthal boundary conditions have the usual form
\begin{eqnarray}
\label{boundary}
\Psi_\alpha(r_e, r_h, \phi_e, \phi_h) &=& \Psi_\alpha(r_e, r_h, \phi_e + 2 \pi, \phi_h), \nonumber \\
\Psi_\alpha(r_e, r_h, \phi_e, \phi_h) &=& \Psi_\alpha(r_e, r_h,
\phi_e, \phi_h + 2 \pi) \, ,
\end{eqnarray}
which leads using Eq.~(\ref{expanG2}) to the following relation
between $l$ and $L$:
\begin{eqnarray}
L \; \mbox{even:} \; l \; \mbox{integer}; \qquad L \; \mbox{odd:}
\; l \; \mbox{half integer} \, .
\end{eqnarray}
The expansion functions $u_{l, L, \alpha}(r_e, r_h)$ obey a
coupled system of Schr\"odinger equations. The confining
potential produces the matrix elements
\begin{eqnarray}
\langle lL | V_a(r_a, \phi_a) |l'L' \rangle = \delta_{l-l',\pm \frac{L-L'}{2}} V^a_{L-L'}(r_a), \nonumber \\
\label{matel} V^a_{k}(r_a) = \frac{1}{2 \pi} \int_0^{2 \pi}
d\phi_a \, V_a(r_a, \phi_a) e^{ik\phi_a} \, ,
\end{eqnarray}
where again $+$ ($-$) sign corresponds to the electron (hole)
From the reality of $V_a(r_a, \phi_a)$ follows $V^a_k =
(V^a_{-k})^*$.
\figb

The confining potential can have an arbitrary symmetry (see
Fig.~\ref{figb} for examples) which is classified by the
two-dimensional point group. There are two types: (i) $C_n$ which
consists of all rotations about the origin by multiples of the
angle $2\pi/n$ and (ii) $D_n$ which adds to the rotations of
$C_n$ reflections with respect to $n$ axes passing through the
origin. Within $C_n$, the potential $V_a(r_a, \phi_a)$ is
invariant under rotations by multiples of the angle $2 \pi / n$.
Changing the integration variable in Eq.~(\ref{matel}) by
$\phi_a' = \phi_a + 2 \pi / n$ we obtain
\begin{eqnarray}
V^a_{k}(r_a) &=& \frac{1}{2 \pi} \int_0^{2 \pi} d\phi_a \,
V_a(r_a, \phi_a) e^{ik(\phi_a + \frac{2 \pi}{n})} \, .
\end{eqnarray}
Comparison with Eq.~(\ref{matel}) shows that
only elements with $k = j n$ ($j$ integer) can be non-zero. Using
similar arguments the reflections in $D_n$ imply $V^a_k =
V^a_{-k}$ to hold. The symmetry properties are summarized in
Tab.~\ref{tabc}.

In the absence of any symmetry ($C_1$) all matrix elements in
Eq.~(\ref{matel}) can be non-zero. For increased symmetry ($C_n$ or
$D_n$, $n > 1$) the matrix decomposes into $n$ block matrices, since the
matrix elements $\langle lL |V_a | l'L' \rangle$ are non-zero
only if $L-L'=n$. In the limiting case of large $n$ the point
groups $C_n$ and $D_n$ converge to the point group of the circle
$O(2)$. Due to the rotational invariance of the circle, one
degree of freedom can be factorized. This is the
total angular motion since the corresponding commutator with the
Hamiltonian vanishes, $[\hat{H}, -i\hbar \partial / \partial
\Phi] = 0$.
\tabc

The Coulomb potential $V_C(r_e, r_h, \phi)$ depends only on the
relative angle, and its matrix elements are diagonal in $L$
\begin{eqnarray}
\langle lL | V_C(r_e, r_h, \phi) | l'L'\rangle = \delta_{LL'} V^C_k(r_e, r_h) \, , \\
V^C_k (r_e, r_h) = \frac{1}{2 \pi} \int_0^{2 \pi} d \phi \,
V_C(r_e, r_h, \phi) \cos(k \phi) \, . \nonumber
\end{eqnarray}
Putting all terms together, the matrix elements of the Hamiltonian are
given by
\begin{eqnarray}
\label{hamexp}
&& \langle lL | \hat{H}(r_e, r_h, \phi, \Phi)| l' L' \rangle = \\
\nonumber && \delta_{ll'} \delta_{LL'} \sum_{a = e,h} \biggl[ \frac{\hbar^2}{2m_a} \biggl\{ -\frac{1}{r_a} \frac{\partial}{\partial r_a} \left (r_a \frac{\partial}{\partial r_a} \right)  \\
&& + \frac{1}{r_a^2} \left ( l \pm \frac{L}{2} + \frac{eB}{2\hbar} r_a^2 \right)^2 \biggr\} \biggr ] - \delta_{LL'} V^C_{l-l'}(r_e, r_h) \nonumber \\
&& + \delta_{l-l', \frac{L-L'}{2}} V_{L-L'}^e(r_e) + \delta_{l'-l, \frac{L-L'}{2}} V_{L-L'}^h(r_h) \,.
\nonumber
\end{eqnarray}

The azimuthal kinetic terms can be rewritten as
\begin{eqnarray}
\label{relfull}
&& \sum_{a = e,h} \frac{\hbar^2}{2m_a r_a^2} \left ( l \pm \frac{L}{2} + \frac{eB}{2 \hbar} r_a^2 \right)^2  =  \nonumber \\
&& \frac{\hbar^2}{2 \mu_X r_X^2} \left (l + \frac{e B}{2 \hbar} r_X^2  + \frac{L}{2} p \right)^2  \\
&+& \frac{\hbar^2}{2 (m_e r_e^2 + m_h r_h^2)} \left (\frac{e B}{2
\hbar} (r_e^2 - r_h^2) + L \right)^2 \nonumber
\end{eqnarray}
with the exciton reduced mass $\mu_X=m_e m_h / (m_e + m_h)$.
Further, $r_X$ is an effective ring radius for the exciton and
$p$ a phase shift, defined as
\begin{eqnarray}
r^2_X = \frac{r_e^2 r_h^2\left(m_e + m_h \right)}{m_e r_e^2 + m_h
r_h^2},  \quad p = \frac{m_h r_h^2 - m_e r_e^2}{m_h r_h^2 + m_e
r_e^2}.
\end{eqnarray}
Equation~(\ref{relfull}) resembles the zero-width ring model
discussed in Ref.~\onlinecite{GGZ2006b}. For $L=0$, the last line
is responsible for the overall increase of energies quadratically
in $B$, while the second line is the source of the X-ABE
oscillations. We will use Eq.~(\ref{relfull}) in Section III for
extracting a reasonable estimate of the $B$ oscillation period.

For dipole-allowed optical interband transitions, the oscillator
strength $f_\alpha$ of the exciton state $\alpha$ is given by the
amplitude of finding electron and hole at the same
place,~\cite{HaKo}
\begin{eqnarray}
\label{os01} f_\alpha = d_{cv} \int d{\mathsf r} \,
\Phi_\alpha({\mathsf r}, {\mathsf r}) \, ,
\end{eqnarray}
where $d_{cv}$ is the interband dipole matrix element.
Introducing the single sublevel approximation and
the expansion Eq.~(\ref{expanG2}), we find
\begin{eqnarray}
\label{os02} f_\alpha =  d_{cv} \sum_l \int_0^\infty dr \, r \,
u_{l, 0, \alpha}(r, r) .
\end{eqnarray}
Only the wave function component with $L = 0$ contributes to the
oscillator strength. States with non-zero oscillator strength are
called bright states, while dark states have $f_\alpha = 0$.
\figc

As an example of a structure with low symmetry, we investigate a
ring with $D_1$ symmetry. Our simple model has two circles as
boundaries. The inner one is centered at the origin (radius
$r_1$), while the outer one (radius $r_2$) is displaced by $b$,
as indicated in Fig.~\ref{figc}. Electron energy spectra have been investigated for such a structure in Ref. \onlinecite{BL2005}, calling the ring {\it eccentric}. We will use in what follows non-circular and eccentric as equivalent names, contrasting to {\it concentric} and circular. The potential is assumed to be
constant with a value $V^a_0$ between these two rings, and set to
zero outside. In the figures, however, zero of energy is set to
the confinement gap of the quantum well made from the ring material.

The matrix element Eq.~(\ref{matel}) reduces to
\begin{eqnarray}
&& \nonumber V_k^a(r_a) = \frac{V^a_0}{2 \pi} \int_{-\pi}^{\pi} d\phi e^{ik\phi} \theta \left ( r_a - r_1 \right ) \times  \\
&& \theta \left ( r_{2}^2 - (r_a \cos (\phi) - b)^2 - r_a^2 \sin^2 (\phi) \right ).
\end{eqnarray}
The unit step functions determine the integration boundaries to
be $\pm \pi$ or $\pm \phi(r_a)$ where
\begin{eqnarray}
\cos(\phi (r_a)) = \frac{1}{2} \frac{b^2 + r_a^2 - r_{2}^2}{r_a
b} \, .
\end{eqnarray}
The result can be given analytically,
\begin{eqnarray} \label{Vanalytic}
V_k^a(r_a) &=& V^a_0 \Bigl [ \delta_{k,0} \theta(r_a - r_1) \theta(r_2 - b - r_a)  \\
&+& \frac{1}{k \pi} \sin(k \phi (r_a)) \theta(r_a - r_2 + b)
\theta(r_2 + b - r_a) \Bigr ], \nonumber
\end{eqnarray}
which properly satisfies the relation $V_k^a = V_{-k}^a$ inherent
to $D_1$ symmetry.

The circular ring is contained as a special case: Setting $b =
0$, only the first line in Eq.~(\ref{Vanalytic}) contributes,
\begin{eqnarray}
V_k^a(r_a) =  V^a_0 \delta_{k,0} \theta(r_2 - r_a) \theta(r_a -
r_1) \, .
\end{eqnarray}
Consequently, the Hamiltonian matrix is fully diagonal in $L$,
and the wave function expansion Eq.~(\ref{expanG2}) reduces to
the sum over $l$, while $L = L_\alpha$ is fixed and therefore a
good quantum number.
\figca

\section{X-ABE of the exciton ground state}

We begin with a discussion of the X-ABE for the exciton ground
state in concentric and eccentric nanorings of type I and II as
schematically plotted in Fig.~\ref{figca}. The material
parameters used are summarized in Tab.~\ref{taba}. The effective
masses are chosen according to the material in which the particle
is found predominantly. The values of the mass density, the sound velocity, and the deformation potentials are taken for the ring material.
\taba

In the calculation, $B$-field strengths up to $B=25\,$T are used which
can be easily achieved in experiment. The radial coordinates in
Eq.~(\ref{hamexp}) have been discretized on a grid of 40 points with a grid step of 0.5 nm. The expansion of the wave function into $l$ and $L$ components has been checked for convergency of the results, leading to a truncation of  $|l|<13$ and $|L|< 10$. The subsequent numerical diagonalization was performed with the improved Lanczos
method~\cite{WS1998} and checked with the Leapfrog method.~\cite{GGZb2005, GGZ2005} 

\subsection{Type I nanoring}

First, a nanoring of GaAs embedded in the surrounding material
Al$_{0.23}$Ga$_{0.77}$As is considered. This type I structure has a
confinement of both electron and hole within the nanoring.
\figd

The dependence of the lowest three exciton energies on magnetic
field is plotted in Fig.~\ref{figd}a. In the case of circular
symmetry ($b=0\,$nm, black curves) the states can be sorted
according to their quantum number $L_\alpha$. The ground state
has $L_\alpha=0$ (solid), the first and the second excited ones
$L_\alpha=1$ (dashed) and $L_\alpha=-1$ (dotted), respectively. The energies of the
first excited states are degenerate at $B=0\,$T. This degeneracy
is lifted at $B \neq 0$ with a relatively small splitting since
the effective radial distance between electron and hole is small.
In order to see this dependence more clearly, let us look upon
Eq.~(\ref{relfull}). Due to the strong radial confinement, the
coordinates $r_a$ in Eq.~(\ref{relfull}) can be replaced by their
expectation values $R_a = \langle r_a \rangle$. Their values
being rather close in the present case ($R_e= 8.0\,$nm, $R_h=
7.8\,$nm), we realize that $L = \pm 1$ gives only a minor
difference in the energy.

The case of non-circular symmetry ($b=1$\,nm, red (grey) curves in
Fig.~\ref{figd}a) is qualitatively not different from the
circular one since the energetically lowest states for different
quantum numbers $L$ do not cross. Their mixing leads to a larger
splitting among states and lifts the degeneracy at $B=0\,$T. More important are the
changes in the oscillator strength (Fig.~\ref{figd}b). The $D_1$
symmetry of the eccentric nanoring implies that there is one
symmetry axis, let us say the $x$-axis, with reflection operator
$\hat{T}_x$. The wave function transforms in the following way
\begin{eqnarray}
\label{symd1}
\hat{T}_x  \Psi_\alpha(x_e, y_e, x_h, y_h) &\equiv& \Psi_\alpha(x_e,-y_e, x_h, -y_h) \nonumber \\
&=& \pm \Psi_\alpha(x_e,y_e, x_h,  y_h).
\end{eqnarray}
With respect to this symmetry operation, all states can be
grouped into even and odd ones. The odd states have zero
oscillator strength which follows immediately from
Eq.~(\ref{symd1}). The doubly degenerate states of the concetric nanoring at $B = 0$~T can form an even and odd linear
combination with respect to $\hat{T}_x$. As the symmetry is
lowered these combinations get mixed. The energetic order of even
and odd states can be estimated for the lowest ones. The ground
state is always even as shown in Fig.~\ref{figd}b. The first
excited is odd since the contribution of the Coulomb interaction
is still large as in the ground state, while the kinetic
energy is lower than for the next even state. Further excited
states have different Coulomb contributions and that is why their
order cannot be determined in general.

All lines in Fig.~\ref{figd}a seem to shift upwards quadratically in $B$, which is indeed the dominant part of the diamagnetic shift. In order to extract the tiny amplitude of the X-ABE oscillations, we have proposed in our previous publication Ref.~\onlinecite{GGZ2006b} to calculate the second derivative of the exciton energy with respect to the $B$-field. This is shown in Fig.~\ref{figd}c. 

The period can be estimated from the second line of Eq.~(\ref{relfull}), assuming that due to the confinement both $\langle r_e \rangle$ and $\langle r_h \rangle$ are almost constant, as 
\begin{eqnarray}
\label{BP} B_P &=& \frac{2 \hbar}{e} \frac{1}{R_X^2},
\end{eqnarray}
which is $B_P = 20.6\,$T in the present case. The minima of the oscillatory component of the energy are found at $B = j B_P$, $j$ being integer, and maxima at $B = j B_P / 2$. Consequently, the minimum of the second derivative is
found at around half of the oscillation period in all cases ($B_P/2 = 10.3\,$T).  

The oscillation amplitude is reduced for the
eccentric nanoring. Here, the non-uniform ring width tends to
push the exciton wave function into the broader part, thus
weakening its ring topology which is a necessary ingredient for
X-ABE oscillations. Note that for $b \geq r_2 - r_1$ the
confinement potential reduces to a banana-shaped quantum dot, and
the ring topology is lost completely.

\subsection{Type II nanoring}

Secondly, we investigate a type II nanoring consisting of InP in
the ring and Ga$_{0.51}$In$_{0.49}$P outside, embedded into an
AlAs barrier along $z$. Strain effects are taken into account
only in so far as the hole is always found inside, $r_h<r_1$. A
full inclusion of strain would modify the exact potential
profile but not the $O(2)$ or $D_1$ symmetry.

For the concetric type II ring, Fig.~\ref{figd}d  illustrates the
crossings among states with different quantum number $L_\alpha$ (black
solid $L_\alpha=0$, dashed $L_\alpha=\pm1$, and dotted $L_\alpha=\pm2$). This crossing
resembles more the ABE for individual carriers and can be traced
back to the much reduced exciton effect in the present type II
nanoring. In order not to overload Fig.~\ref{figd}d, some higher states have been omitted. If the symmetry is reduced to $D_1$ (red [grey] curves), all
states are mixed similar to the type I nanoring. Consequently,
all crossings become anti-crossings and the energy
dependence on the magnetic field differs not much from the type I nanoring, with reduced X-ABE oscillations
(Fig.~\ref{figd}f).

The difference between type I and II nanorings becomes apparent
when the oscillator strength is studied as depicted in
Fig.~\ref{figd}e. The oscillator strength corresponding to different exciton states changes with the magnetic field as the character of the exciton wave function changes itself: The main component of the ground state wave function is
$L=0$ at $B=0\,$T and shifts to $L=1$ at $B=10\,$T. This results in
a decrease of the oscillator strength which is transferred to the
first excited state (and increased due to the larger
electron-hole overlap). Later on it is transferred to the second
and higher excited states where different quantum numbers $L$ mix
strongly. Even though there is not any strict rule for the oscillator strength conservation, Fig.~\ref{figd}e suggests that it is approximately valid for the lowest exciton states. The rapid decay of the ground state oscillator strength with increasing magnetic field has interesting implications for the exciton kinetics (Section IV).

Unlike the type I nanoring, the oscillation period changes with
$b$ (Fig.~\ref{figd}f). This is also accompanied by a decrease of
the oscillation amplitude due to the loss of the ring topology
with increasing $b$. The oscillation period is now determined by the periodic change of the ground state main component from $L$ to $L+1$, thus by the center-of-mass motion. The exciton relative motion with the period $B_P$ in Eq.~(\ref{BP}) plays only a minor role for {\it every} value of $b \neq 0$. The value of the center-of-mass period can be estimated from the last term of Eq.~(\ref{relfull}) assuming again strong radial confinement,
\begin{eqnarray}
\label{per2}
B_{P, 2} = \frac{2 \hbar}{e} \frac{1}{R_e^2 - R_h^2}.
\end{eqnarray}
In the present case, taking $R_e = \langle r_e \rangle = 12.3\,$nm
and $R_h = \langle r_h \rangle = 4.4\,$nm we obtain $B_{P, 2} =
9.9\,$T, which agrees well with the observed period in
Fig.~\ref{figd}f.
\figf

In Fig.~\ref{figf} correlated one-particle densities are plotted.
They are defined as conditional probability to find a particle,
either electron or hole, in the exciton while fixing the coordinates of the other
particle at a certain position marked by a large dot. For the
formal definition, see Eqs.~(\ref{densitye}, \ref{densityh}). This
concept has been used also for analyzing the two-electron Wigner
molecule in a quantum dot.\cite{PSN2004} An inspection of Fig.~\ref{figf} reveals how the loss of the
circular symmetry in the eccentric nanoring modifies the wave
function. By fixing the hole position at the right (wide) side of
the ring, most of the electron density is found there as well
(Fig.~\ref{figf}a), which resembles the situation in a concentric
ring (not shown). The same is true for the hole
(Fig.~\ref{figf}c). On the other hand, by fixing either the
electron or the hole on the opposite narrow side of the ring, the
picture changes. Due to the strength of the ring confinement and
the weakness of the Coulomb interaction in type II structures,
the electron is still found on the right where the confinement
energy is minimal (Fig.~\ref{figf}b). Since the electron density
is very small on the opposite side, the Coulomb correlation seen
by the hole is tiny, and an almost circular -- i.e. one-particle
like -- hole density is found in Fig.~\ref{figf}d.

\figi

Finally, let us briefly consider the case of a free electron-hole pair which has been investigated for type I nanorings in Ref. \onlinecite{BL2005}. Formally, we switch off the Coulomb interaction $V_C$ in the Hamiltonian Eq.(\ref{H01}). The free electron-hole transitions shown as dashed curves in Fig.~\ref{figi}a are dominated by the electron level since the hole part refers to an almost fixed angular quantum number due to the small effective hole radius of 4 nm. Therefore, the first levels can be characterized by the electron quantum numbers $l_e = 0,-1,+1$. Their degeneracy at zero magnetic field is lifted when switching on the Coulomb interaction (solid curves), similarly to Fig.~\ref{figd}d. The overall down shift in energy is a measure of the exciton binding energy. 

The second derivative displayed in Fig.~\ref{figi}b shows that the diamagnetic shift (being proportional to the relative distance) of the free electron-hole pair is larger than in the exciton. The reason is the Coulomb attraction in the exciton which brings electron and hole closer together. The oscillation amplitude, however, is getting stronger for the exciton. We conclude that the ring topology of the electron part is stabilized due to the Coulomb attraction with the hole. Therefore, surprisingly in the specific type II example, the ring eccentricity is felt not as strong for the exciton compared with the free e-h- pair.”

To summarize this section, we have demonstrated that the amplitude of X-ABE weakens when
going from circular to non-circular symmetry since the wave
function tends to lose its ring topology. However, if the
asymmetry is weak and electron and hole are spatially separated
as in type II structures, the ground state still exhibits oscillations in its energy, while being always bright and
observable by optical means.

\section{Kinetic equations}

For calculating the photoluminescence emitted from the
nanoring, we need to know the occupation $N_\alpha$ of each
exciton states. For the linear density regime and incoherent
excitation, the relevant set of kinetic equations has been
derived in Ref.~\onlinecite{ZR1997},
\begin{eqnarray}
\label{kinet} \frac{d N_\alpha}{dt} = g_\alpha + \sum_\beta
\gamma_{\alpha \beta} N_\beta - ( r_\alpha + d_\alpha +
\sum_\beta \gamma_{\beta \alpha} ) N_\alpha \, .~~
\end{eqnarray}
Here, $g_\alpha$ is a state dependent generation term which
stands for the last term in a chain of optical phonon emission
events after optical interband excitation.\cite{SAK2001} The
radiative decay rate $r_\alpha$ of a localized exciton state
contains emission into both TE and TM polarization and is found
proportional to the squared oscillator strength,~\cite{Runge2002}
\begin{eqnarray}
\label{radiv0}
r_{\alpha} = \frac{4}{3}\frac{E_g^3 n_R}{\hbar^4 c^3} |f_\alpha|^2 \, ,
\end{eqnarray}
where $n_R$ is the refractive index and $E_g$ the band gap. In
Eq.~(\ref{kinet}), $d_\alpha$ is a phenomenological non-radiative
decay rate, representing processes as e. g. exciton annihilation
via impurities, escape into the wetting layer, or Auger
processes. These processes are not treated explicitly in the present theory.

The acoustic phonon scattering rates $\gamma_{\alpha
\beta}$ are defined as
\begin{eqnarray}
\label{gamma}
\nonumber \gamma_{\alpha \beta} &=& \frac{2 \pi}{\hbar} \sum_{{\mathsf q}} |t^{{\mathsf q}}_{\alpha \beta}|^2
\left[ (n_B(\hbar \omega_q) + 1)
    \delta(E_\beta - E_\alpha - \hbar \omega_q ) \right. \\
&& + \left.  n_B(\hbar \omega_q) \delta(E_\beta - E_\alpha + \hbar \omega_q ) \right] \,
\end{eqnarray}
where  $n_B(\hbar \omega_q)$ is the Bose-Einstein distribution of
acoustic phonons with dispersion $\hbar \omega_{q} = \hbar s q$
($s$ -- sound velocity). They obey the relation of detailed
balance between in- and out scattering of a given state,
\begin{eqnarray} \label{detailed}
\gamma_{\beta \alpha} = \gamma_{\alpha \beta} \, e^{(E_\alpha -
E_\beta)/k_B T}\, ,
\end{eqnarray}
with the phonon (i.e. lattice) temperature $T$. Strictly
speaking, lattice vibrations in nanostructures differ from the
respective bulk ones,\cite{StDu} but for the present purpose this refinement is of minor importance.\cite{GZ2007} The evaluation of the exciton-phonon matrix elements $t^{{\mathsf
q}}_{\alpha \beta}$ for deformation potential scattering is given
in the Appendix. 

In the exciton scattering with acoustic phonons we have taken into account only the spin diagonal part. Indeed, there is non-diagonal scattering ("spin-flip") as well due to the lack of the inversion symmetry in nanostructures -- Rashba effect.\cite{ALS2002} Under linearly polarized excitation, the initial generation $g_\alpha$ is spin independent. Assuming that in the final stage of thermalization spin-flip processes can be neglected, we expect an equal occupation of spin up and down bright states. This is supported by recent experiments showing almost no difference in intensities between $\sigma^+$ and $\sigma^-$ polarized lines.\cite{SP2002}

The kinetic equations are solved numerically for
the steady state situation $dN_\alpha / dt = 0$. The resulting
occupations $N_\alpha$ enter the photoluminescence spectrum
$I(E)$,
\begin{eqnarray}
\label{pl} I(E) &=& \sum_\alpha r_\alpha \,N_\alpha \, \delta(E -
E_\alpha) \, .
\end{eqnarray}
Note that the linear absorption spectrum $D(E)$ does not depend
on the occupations,
\begin{eqnarray} \label{abs} D(E) &=& \sum_\alpha
r_\alpha \,\delta(E - E_\alpha) \, .
\end{eqnarray}
Again, a constant prefactor has been omitted.

Summing Eq.~(\ref{kinet}) over all states $\alpha$, the phonon
scattering terms cancel, and the following conservation law for
the total pump rate $P$ is found
\begin{eqnarray}
\label{law} P \equiv \sum_\alpha g_\alpha = \sum_\alpha (r_\alpha
+ d_\alpha) N_\alpha \, .
\end{eqnarray}
We will exploit this relation in Section V in the discussion of
PL quenching. i.e. a decrease of the PL intensity with $B$-field.

\section{Absorption and Photoluminescence}

Since in type I nanorings, the exciton energies and oscillator
strengths do not depend much on the magnetic field, interesting
effects like PL quenching are not to be expected. Therefore, we
focus here on a type II system and investigate InP/GaInP nanorings with radii ~$r_1 = 8\,$nm, $r_2 =
16\,$nm. Results for an eccentric nanoring (c) are compared with a sligthly eccentric one (n), having a center displacement $b$ of $0.5$\,nm as used in Fig.~\ref{figd}, right panel. The radiative rates of the lowest state $\alpha=0$ are rather small ($r^c_{0} = 0.042\,\mbox{ns}^{-1}$ and $r^n_{0} = 0.027\,\mbox{ns}^{-1}$ at $B=0\,$T)
due to the tiny overlap between electron and hole in the wave
function, as specific for any type II structure.  Similar small
radiative rates have been calculated for spatially indirect
excitons in coupled quantum wells.\cite{Zim2006} The
exciton-phonon scattering rates from the first excited state down
to the ground state are $\gamma_{01}^c = 102\,\mbox{ns}^{-1}$ and $\gamma_{01}^n = 138\,\mbox{ns}^{-1}$ at
$B=0\,$T. These numbers clearly indicate that the exciton-phonon
scattering dominates the kinetics.

\figga

The absorption spectra of the circular and non-circular nanorings plotted in Fig.~\ref{figga} are calculated for linearly polarized light where both spin components Eq.~(\ref{spins}) with $M=\pm1$ are present. Therefore, all lines appear as Zeeman splitted doublets.  

The absorption spectra plotted in Fig.~\ref{figga} show pronounced differences between both nanorings: In the concetric nanoring only two doublets having quantum number $L_\alpha = 0$ are visible (Fig.~\ref{figga}a). Above $B=5$\,T, the lower state is not any longer the ground state (compare Fig. \ref{figd}d). In the eccentric ring the oscillator strength is transferred from the ground state to higher states with increasing magnetic field (Fig.~\ref{figga}b) as already discussed in detail in Section III--B.  Although the oscillation of the lowest bright state  can be hardly seen in both cases, their second derivatives reveals them clearly (see Fig.~\ref{figd}f). The excited states exhibit much stronger oscillations since here the Coulomb attraction between electron and hole acts much less. Both periods agree well with Eqs.~(\ref{BP}) and~(\ref{per2}), respectively.

\figgb

In view of an easier interpretation of the calculated spectra, we envisage in Figs.~\ref{figgb} and~\ref{figgc} a PL detection with circularly polarized light $\sigma^+$. Thus, only the lower Zeeman splitted lines are seen. For solving the kinetic equations, nine exciton states were taken into account. First, we investigate the case without non-radiative decay, $d_\alpha=0$ (Fig.~\ref{figgb}). Due to the large energy difference between the first and the second bright state, only the lowest one is visible for both temperatures (Fig.~\ref{figgb}a and c). The results for the eccentric ring are not so simple. Although the oscillator strength of the ground state goes down appreciably (at $B=25\,$T it is less than $10^{-2}$ of its value at $B=0\,$T), the luminescence line at $T = 4$\,K (Fig.~\ref{figgb}b) has almost constant intensity, and no line quenching is seen. This is a consequence of the conservation law Eq.~(\ref{law}) and will be discussed below in more detail. At elevated temperature, other lines are seen as well, but their intensities have a different $B$ dependence compared to the absorption, which signals the role of exciton 
occupation (Fig.~\ref{figgb}d).
\figgc

Second, non-radiative decay is included in Fig~\ref{figgc}. For the discussion, three rates respectivelly times are important: (i) the radiative rate $r_\alpha$ (tens of nanoseconds), (ii) the exciton-phonon scattering rate $\gamma_{\alpha \beta}$ (several picoseconds), and (iii) the non-radiative rate $d_\alpha$ which is a phenomenological input here. With the assumption $d_\alpha = 1/(100\,$ps) we use a value which dominates over the extremely small radiative rate, but is well below the phonon scattering rate. We start again with the discussion of the circular ring where the situation is simple. Due to the change of the ground state from bright to dark one at  $B>5$\,T, these excitons decay predominantly non-radiatively, which results in a distinct line quenching (Fig.~\ref{figgc}a and c). On the other hand, for the non-circular nanoring more lines are seen in the PL, in particular at low temperatures (Figs.~\ref{figgc}b and d). Obviously, complete equilibration with the lattice temperature is no longer reached. For the unrealistic case of non-radiative rates being stronger than the phonon scattering, the exciton occupation is simply given by the ratio between pump rate $g_\alpha$ and decay $d_\alpha$. Keeping both quantities constant, the exciton occupation gets constant, and the PL spectra would coincide with the absorption spectra. Although this extreme limit is not reached with the actual parameters, the tendency is clearly observable in Fig.~\ref{figgc}.
\figh

Let us now concentrate on the integrated photoluminescence $I$
which follows from Eq.~(\ref{pl}) as
\begin{eqnarray}
\label{totalpl} I \equiv \int d\omega \, I(\omega) =
\sum_\alpha r_\alpha N_\alpha \, .
\end{eqnarray}
Its relation to the pump rate $P$ will be called photoluminescence yield $Y$,
\begin{eqnarray}
\label{full} Y \equiv \frac{I}{P} = \frac{\sum_\alpha r_\alpha
N_\alpha}{\sum_\alpha g_\alpha} \, .
\end{eqnarray}
Without non-radiative decay ($d_\alpha = 0$), the conservation
law Eq.~(\ref{law}) gives immediately $Y = 1$ independent of $B$
-- each excited exciton (or in general electron--hole pair) decays into
one emitted photon (full curve in Fig.~\ref{figh}a and b).

Things change a lot if non-radiative decay channels are
included, $d_\alpha > 0$. Due to the extremely small radiative
rates, these processes can even dominate the exciton decay,
leading to a yield much below unity (Fig.~\ref{figh}c and d). Since the ground state of the concentric ring is getting dark at around $B=5$\,T, a steep decay of the yield follows. In the eccentric nanoring the yield goes down not as abruptly since the oscillator strength of the ground state decays more slowly. The slight oscillations seen in the full curve of Fig.~\ref{figh}d are related to the changing
level distances which influence the individual phonon scattering
rates.  We conclude that the quenching of the total PL is
intimately related to non-radiative processes. The quenching is
not as dramatic at elevated temperatures since higher exciton
states contribute more and more to the total emission (not
shown).

In all cases studied here, the phonon scattering rates are
dominant. Therefore, it can be expected that the occupations of
the different exciton states deviate not too much from
equilibrium which in the present low-density case, in accordance
with Eq.~(\ref{detailed}), is
characterized by the Maxwell-Boltzmann (MB) distribution
\begin{eqnarray}
\label{MB}
 N_\alpha = C \exp(-\beta E_\alpha)
\end{eqnarray}
with $\beta = 1 / k_B T$. Within this approximation, the PL
yield is given by
\begin{eqnarray}
\label{YMB}
 Y_\mathrm{MB} = \frac{\sum_\alpha r_\alpha e^{-\beta E_\alpha}}{\sum_\alpha (r_\alpha + d_\alpha) e^{-\beta E_\alpha}}
\end{eqnarray}
and shown as dotted curves in Fig.~\ref{figh}, indeed not very
much different from the full calculation. This is the right place
to discuss Ref.~\onlinecite{SUG2004} which was the first attempt
to calculate the exciton PL for a nanoring. Concerning the
exciton kinetics, they (i) have assumed a Maxwell-Boltzmann
distribution for the excitons and (ii) normalized the integrated
PL to constant exciton density $N = \sum_\alpha N_\alpha$ --
and not to pump rate $P$ -- resulting in
\begin{eqnarray}
\label{govorov} Y_\mathrm{MBN} = \frac{\sum_\alpha r_\alpha
e^{-\beta E_\alpha }}{\sum_\alpha e^{-\beta E_\alpha}} \, .
\end{eqnarray}
This normalization makes a pronounced difference when
non-radiative processes are absent (dashed curve in
Fig.~\ref{figh}a), but gives nearly identical results with
Eq.~(\ref{YMB}) when these dominate (dashed curve in
Fig.~\ref{figh}b). $Y_\mathrm{MBN}$ has been normalized to
$Y_\mathrm{MB}$ at $B=0\,$T.

While this discussion refers to PL {\it quenching}, in
Ref.~\onlinecite{SUG2004} PL {\it blinking} was proposed, too:
The integrated PL goes down and up in dependence on $B$, since
the ground state switches between bright and almost dark
behavior. The authors have used a ring model with zero width and
rather close radii for electron an hole. The latter seems to be decisive for the blinking effect to occur. For the type II nanoring studied in the
present work, such a blinking cannot be expected since the
average electron and hole radii are rather different
($R_e=12.1\,$nm vs. $R_h=4.4\,$nm). Moreover, the
blinking effect predicted in Ref.~\onlinecite{SUG2004} has been questioned recently by noting that
a better account of the Coulomb interaction is
needed.\cite{BPS+2006}

\section{Conclusions and outlook}

We have derived the Hamiltonian for excitons in nanorings with
{\it finite width} and {\it arbitrary symmetry}. Sorting its matrix elements
according to the symmetry point group of the confining potential
we have shown how the different wave function components are
coupled. Two prototype systems of type I (GaAs/AlGaAs) and type
II (InP/GaInP) have been considered. The numerical investigations
showed that: (i) The oscillation amplitude of the exciton
Aharonov-Bohm effect decreases when going from circular to
non-circular symmetry due to the additional localization of the
exciton. Moreover, the oscillation period in type II nanorings
changes from relative-motion induced to being determined by
periodic changes of the center-of-mass wave function. (ii) The
exciton ground state in the non-circular (eccentric) nanoring remains always
optically active since the total angular momentum is no longer a
good quantum number. However, its oscillator strength can be
extremely small in type II nanorings.

Further, we have investigated the {\it exciton kinetics} within
a model which includes acoustic phonon scattering and radiative
and non-radiative decay. Our study of a {\it slightly}
eccentric type II nanoring has revealed that the oscillations
of the excited states are clearly visible and that the amplitude of the ground state
oscillations is weaker than in the concentric ring.
Calculating temperature dependent occupations of exciton states
we could show how the appearance and disappearance of single
lines in PL are related to their counterparts in absorption. The
presence of non-radiative decay is decisive for the quenching of
the integrated photoluminescence. If the exciton-phonon
scattering dominates over all decay rates, a simplified
description using a Maxwell-Boltzmann exciton distribution works
reasonably well.

One unexpected finding is that a small violation of the circular
symmetry improves the observability of the X-ABE since here all
states are optically active, and anticrossings periodic with $B$
can be easily seen in the second derivative of the ground state
energy. Non-radiative decay channels allow to see more lines in
PL, thus improving the observability of X-ABE for excited states
even at low temperatures. Therefore, a slightly asymmetric
nanoring of type II seems to be the best candidate for the
experimental confirmation of the exciton Aharonov-Bohm effect in
linear optics, i.e. absorption or photoluminescence. From a
practical point of view, an exactly circular nanoring would be
rather the exception than the rule, given the uncertainties of
nanostructure growth on patterned substrate, not to think of
self-organized ring formation.

Quite recently, the possibility of implementing a flux qubit in
small non-superconducting rings has been
discussed.~\cite{ZKS+2006} Since a persistent current due to
excitons can be initiated and controlled
optically,~\cite{GGZ2006b} it can be speculated that an exciton
qubit in a nanoring can be formed whose function rests upon the
exciton Aharonov-Bohm effect.

\section{Acknowledgement}
We acknowledge fruitful discussions with and helpful comments of
L. Wendler and E. A. Muljarov. M. G. acknowledges financial
support from the Graduate school Nr. 1025 of the Deutsche
Forschungsgemeinschaft.

\appendix

\section{Technical details}

The correlated one-particle densities are defined as conditional probabilities
\begin{eqnarray}
\label{densitye} n^{(e)}_\alpha(r_{e},\phi_e)= \left |
\Psi_\alpha \left (r_e, r, \phi = \phi_e - \varphi, \Phi =
\frac{\phi_e + \varphi}{2} \right ) \right |^2 \nonumber \\
\\
\label{densityh} 
n^{(h)}_\alpha(r_{h},\phi_h)= \left | \Psi_\alpha \left (r, r_h,
\phi = \varphi-\phi_h, \Phi = \frac{\phi_h + \varphi}{2} \right )
\right |^2 \nonumber \\
\end{eqnarray}
where the coordinates of the other particle are fixed at $(r,
\varphi)$. Note that this definition differs from the previous one in Ref.~\onlinecite{GGZ2006b} where an additional integration over $r$ was performed.

The averaged Coulomb potential is for convenience approximated by\cite{Z1986}
\begin{eqnarray}
\nonumber
V_C(r_e,r_h,\phi) = -  \frac{e^2}{4\pi \epsilon_0 \epsilon_S}
\frac{1}{b_C L_z} \mbox{arcsinh}\left(\frac{b_C L_z}{r} \right),
\end{eqnarray}
where $r= \sqrt{r_e^2 + r_h^2 - 2 r_e r_h \cos(\phi)}$, and $L_z$
is the width of the quantum well (4\,nm in our case). $b_C = 0.6$
is an effective parameter which has been fitted to give
reasonable agreement with the quantum well Coulomb potential.

The phonon matrix elements are defined as\cite{T1985}
\begin{eqnarray}
t^{{\mathsf q}}_{\alpha \beta} &=& \sqrt{\frac{\hbar \omega_q}{2 s^2 \rho_M V}} \int
    \int d{\mathsf r}_e \; d{\mathsf r}_h \; \Phi^*_\alpha({\mathsf r}_e, {\mathsf r}_h) \times  \nonumber \\
&& \Bigl (D_c \exp(i{\mathsf q} {\mathsf r}_e) - D_v  \exp(i{\mathsf q}{\mathsf r}_h) \Bigr) \Phi_\beta({\mathsf r}_e, {\mathsf r}_h),\nonumber \\
\end{eqnarray}
where $\rho_M$ is the mass density, $V$ the sample volume, and
$D_c$ ($D_v$) the deformation potential for electron (hole).
Using the single sublevel approximation and the
expansion of the wave function Eq.~(\ref{expanG2}), we obtain in
cylindrical coordinates for the three-dimensional momentum ${\mathsf q}$
\begin{eqnarray}
&& t^{{\mathsf q}}_{\alpha \beta} = \sqrt{\frac{\hbar \omega_q}{2 s^2 \rho_M V}} \sum_{L, L'} e^{-i (L - L') \phi_q} \times \nonumber \\
&& \left (  S^{\alpha \beta e}_{L, L'} (q_\|) K_e(q_z) D_c -  S^{\alpha \beta h}_{L, L'} (q_\|)K_h(q_z) D_v \right), \nonumber \\
\end{eqnarray}
introducing the state dependent overlap functions $S^{\alpha
\beta e(h)}_{L, L'} (q_\|)$ and a $z$-dependent contribution
\begin{eqnarray}
\label{overl1}
&& S^{\alpha \beta e (h) }_{L, L'} (q_\|) = \sum_l \int_0^\infty dr_e \, r_e \, dr_h \, r_h \, u_{l, L, \alpha}(r_e, r_h) \nonumber \\
&& \times u_{l \mp \frac{L-L'}{2}, L', \beta}(r_e, r_h) \, J_{L-L'} (q_\| r_{e(h)}) ; \\
&& K_a(q_z) = \int dz \, v_a^2(z) e^{- i q_z z} \, .
\end{eqnarray}
$J_{L-L'}(x)$ are Bessel functions of the first kind, $v_a(z)$ is the confinement wave function in $z$-direction, and the
symmetry $S^{\alpha \beta e(h)}_{L,L'} (q_\|) = S^{\beta \alpha
e(h)}_{L',L}(q_\|)$ holds. In order to get the scattering rate
$\gamma_{\alpha \beta}$, the matrix element $t^{{\mathsf q}}_{\alpha
\beta}$ squared has to be integrated over ${\mathsf q}$. The
integration over $q_z$ is performed using the energy conserving
delta function in Eq.~\ref{gamma} which gives $|\Delta_{\alpha\beta}| = \hbar
sq$ with the energy difference between states $\Delta_{\alpha\beta} =
E_\alpha - E_\beta$. Defining the overlap sum
\begin{eqnarray}
\quad W^{ab}_{\alpha \beta}(q_\|) &=& \sum_{L M N} S^{\alpha \beta a}_{L,L-N} (q_\|) S^{\alpha \beta b}_{M,M-N} (q_\|),
\end{eqnarray}
the final expression is obtained as
\begin{widetext}
\begin{eqnarray}
\label{general}
\nonumber \gamma_{\alpha \beta} = \frac{n_B(\Delta_{\alpha\beta}) \Delta_{\alpha\beta}}{2 \pi \hbar^2 s^3 \rho_M}  \int_0^q \frac{dq_\| \, q_\|}{\sqrt{1 - q^2_\|/q^2}}
\left [ W^{ee}_{\alpha \beta}(q_\|)  K^2_e(q_z) D_c^2 + W^{hh}_{\alpha \beta}(q_\| ) K^2_h(q_z) D_v^2 - 2 W^{eh}_{\alpha \beta}(q_\| ) K_e(q_z) K_h(q_z) D_c D_v \right], \\
\end{eqnarray}
\end{widetext}
where $q_z = \sqrt{q^2 - q_\|^2}$. Note that both phonon emission
and absorption processes (first and second term in Eq.~(\ref{gamma}),
resp.) are included here, since $\Delta_{\alpha\beta}$ can
have both signs.

In order to simplify further, an approximation introduced in
Ref.~\onlinecite{ZR1997} is adopted here, too. The strongest
confinement is found in the growth direction $z$ which allows to
put $K^2_a(q_z) \sim K^2_a(0) = 1$. Further, a rapid decay of
$W^{ab}_{\alpha \beta}(q_\|)$ is assumed, well before the
integration limit $q$ is reached. Then, the integral in
Eq.~(\ref{general}) can be approximated by
\begin{eqnarray}
\int_0^q \frac{dq_\| \, q_\|}{\sqrt{1 - q^2_\|/q^2}}\,[\cdots ]
\approx \int_0^\infty dq_\| \, q_\| \,[ \cdots] \,, \nonumber
\end{eqnarray}
which allows to integrate over $q_\|$ analytically using \cite{AS1972}
\begin{eqnarray}
\int_0^\infty  dq_\| \, q_\| \, J_N(r_a q_\|) \, J_{N}(r_b q_\|)
= \frac{1}{r_a} \delta(r_a - r_b) \, .\nonumber
\end{eqnarray}
The general expression Eq.~(\ref{general}) reduces to
\begin{eqnarray}
\label{approx}
\gamma_{\alpha \beta} &=& \frac{n_B(\Delta_{\alpha\beta}) \Delta_{\alpha\beta}}{2 \pi \hbar^2 s^3 \rho_M}  \times \\
&& \left [ X^{ee}_{\alpha \beta} D_c^2 + X^{hh}_{\alpha \beta}
D_v^2 - 2 X^{eh}_{\alpha \beta} D_c D_v \right] \, , \nonumber
\end{eqnarray}
introducing the following abbreviations:
\begin{eqnarray}
\nonumber X^{ab}_{\alpha \beta} &=& \sum_{L M N} \int_0^\infty dr \, r \,
\xi^{\alpha \beta a}_{L L - N}(r) \xi^{\alpha \beta b}_{M M - N}(r)\,, \\
\nonumber \xi^{\alpha \beta e}_{L L'}(r) &=& \sum_l \int_0^\infty dr' \, r' \,
u_{l, L, \alpha}(r, r') u_{l - \frac{L-L'}{2}, L', \beta}(r, r') \, , \\
\nonumber \xi^{\alpha \beta h}_{L L'}(r) &=& \sum_l \int_0^\infty dr' \, r' \,
u_{l, L, \alpha}(r', r) u_{l + \frac{L-L'}{2}, L', \beta}(r', r) \, . \\
\nonumber
\end{eqnarray}
The detailed balance Eq.~(\ref{detailed}) can be checked in
Eq.~(\ref{approx}) quite directly.

\bibliographystyle{apsrev}

\end{document}